# The role of defects in the etching of graphene by intercalated oxygen

Tianbai Li and Jory A. Yarmoff*

*Department of Physics and Astronomy, University of California, Riverside, Riverside CA 92521*

**Abstract**

Graphene is one of the most promising 2D materials for various applications due to its unique electronic properties and high thermal stability. In previous studies, it was shown that when graphene is deposited onto some transition metal substrates, small molecules, such as $O_2$, intercalate between the graphene and the substrate and react to partially etch the graphene film when heated to desorb the intercalates. Here, carbon vacancy defects are intentionally formed on Gr/Ru(0001) and their effect on the intercalation of oxygen and etching of the graphene layer are investigated. 50 eV $Ar^+$ sputtering with a low fluence is used to create isolated single vacancy defects in the graphene overlayer and helium low energy ion scattering (LEIS) is employed for surface analysis. It is found that the defects both ease the intercalation of the oxygen molecules and improve the etching efficiency of the graphene during annealing.

---

*Corresponding author, E-mail: yarmoff@ucr.edu

# I. Introduction

Graphene (Gr), an atomically thin sheet of $sp^2$ bonded carbon, is regarded as a promising material for future carbon-based device architectures due to its particular electronic properties [1-3]. The interface of graphene with other materials is a key part of developing large-scale graphene growth and forming conventional device contacts [4-6]. Moreover, graphene can also be used as a protection layer on metal surfaces to avoid oxidation or corrosion due to its chemical and thermal stability [7,8]. It is reported, however, that small molecules, such as $O_2$, CO and $H_2O$, intercalate between a deposited graphene overlayer and the metal substrate when exposed above room temperature or in the presence of defects [9-12].

In previous work [13], helium (He) low energy ion scattering (LEIS) was used to investigate the intercalation of oxygen underneath a Gr film deposited onto a Ru(0001) substrate. It was shown that the $O_2$ intercalates when exposed at 650 K and that it remains in molecular form. It was further shown that the intercalated $O_2$ desorbs from Gr/Ru(0001) when it is annealed to 800 K and, in doing so, etches away some of the graphene. The intercalation process extends progressively from the edge of the graphene layer towards the center and a post-annealing at 823 K leads to a rapid etching of the graphene edge, as shown in Ref. [10], which implies that the edge of the graphene film is more reactive than the intact areas of the film. Note that calculations have also suggested the atoms in the vicinity of a carbon vacancy defect also have a high reactivity [14]. It is thus important to study how the defects in deposited graphene layers, which commonly occur during fabrication and cleaning procedures, affect the intercalation of small molecules and the consequent etching reaction when the material is heated to remove the intercalates. In this paper, small defects are induced by a low fluence $Ar^+$ sputtering of Gr/Ru(0001), and the results show that the defects facilitate both the intercalation of oxygen and the etching of the graphene during the thermal desorption.

The major tool used in this experiment is LEIS, which is well known for its extremely high surface sensitivity [15], and LEIS has been previously applied to measure impurities in Gr overlayers [16]. The surface sensitivity is high when detecting the projectiles that scatter as ions because the Auger neutralization of $He^+$ ions during the scattering process causes most of the ions that experience multiple scattering to be neutralized. By adjusting the scattering angle, a configuration can be obtained that probes only the Gr layer and any adsorbates, or the Gr layer along with any intercalates and revealed areas of the substrate.

## II. Experimental procedure

The experiment was conducted in an ultra-high vacuum (UHV) chamber that has a base pressure of $4\times10^{-10}$ Torr and is equipped with an ion bombardment gun (Varian) for sputtering, low energy electron diffraction (LEED) optics (Varian) and the equipment needed for the LEIS experiments, as described below. The 1 cm diameter sample is mounted on the foot of a manipulator that allows for rotations about both the polar and azimuthal angles. The manipulator foot contains an e-beam heater that uses a rhenium-coated tungsten filament that is floated at a high negative voltage to heat the sample up to 1400 K. The temperature is measured by K-type thermocouples spot welded to the sample plate.

A standard procedure is used for the preparation of a clean and well-ordered Ru(0001) surface, as reported in the literature [17,18]. The approach includes both ion bombardment and annealing (IBA) and a chemical reaction. First, a 30 min 500 eV $Ar^+$ ion sputtering is performed using a flux of $4\times10^{13}$ ions $sec^{-1}$ $cm^{-2}$. The spot size of the sputtering beam is $3\times3$ $cm^2$, as measured by a Faraday cup mounted near the sample holder, to ensure that the entire sample is uniformly bombarded. After sputtering, the sample is annealed under $4\times10^{-8}$ Torr of $O_2$ at 1100 K for 8 min followed by a flash annealing at 1300 K for 2 min under UHV to remove the carbon contaminants via chemical reaction with oxygen, as well as any oxygen residue. To acquire a sufficiently clean and well-ordered surface, this process is normally repeated several times. The cleanliness and crystallinity of the sample surface is verified with LEIS and LEED.

The graphene layer is prepared by a chemical vapor deposition (CVD) method as reported in the literature [18]. This involves annealing the Ru(0001) sample to 900 K under $1.5\times10^{-7}$ Torr of ethylene for 5 min, followed by a flash annealing under vacuum at 1200 K and then cooling to 450 K for another 5 min. Repeating this growth cycle 3 or 4 times is generally required to get a fully-covered, single and continuous graphene layer on the Ru(0001) surface. The quality of the graphene layer is monitored with LEIS and LEED.

Carbon vacancy defects are produced via $Ar^+$ sputtering performed on the as-prepared Gr/Ru(0001) sample, as detailed in Ref. [14]. To remove only a few carbon atoms from Gr, a light sputtering is performed by defocusing the beam and setting the ion energy to 50 eV. As measured by the Faraday cup, the average beam density is 10 pA $mm^{-2}$ and two different sputtering times (1 min and 3 min) are employed. The total fluences of the 1 min and 3 min sputtering treatments are $4\times10^{11}$ ions $cm^{-2}$ and $1.2\times10^{12}$ ions $cm^{-2}$, respectively. The lattice constant for free-standing graphene at room temperature is about 2.45 Å [19] so that there are about $2\times10^{14}$ C atoms within each $cm^2$. As a good approximation, only 1 out of 167 Gr carbon atoms is impacted during the 3 min light sputtering. Considering the low energy and flux of the $Ar^+$ ions, it is assumed that such a light sputtering leads to randomly distributed and isolated single carbon vacancies.

After introduction of defects in Gr/Ru(0001), $O_2$ exposures are performed at a pressure of $1.5\times10^{-6}$ Torr with the sample held at 650 K to produce intercalated oxygen. The location of the oxygen beneath the Gr layer is confirmed with 3 keV $He^+$ ion scattering using the method reported in Ref. [13]. Exposures in this paper are reported in units of Langmiurs (L), where 1 L = $1\times10^{-6}$ torr sec. The sample is held at 650 K until the $O_2$ is evacuated from the chamber.

A differentially pumped ion gun (PHI model 04-303) is used for the $He^+$ LEIS measurements. The incident $He^+$ ion beam has a diameter of 1.6 mm and a total sample current of 1.5 nA. The scattered ions are collected by a Comstock AC-901 hemispherical electrostatic analyzer (ESA) that is mounted on the rotatable platform inside the UHV chamber, which allows for the scattering angle to be adjusted. The data collected in this paper uses a specular geometry in which the incident and outgoing angles are always equal with respect to the surface normal as the scattering angle is varied.

To test for the occurrence of any beam-induced surface damage, five spectra were collected successively from the same spot of a Gr-covered Ru sample, and it was found that the C LEIS peak intensity did not change. To completely avoid beam damage, the samples used in the present experiments are re-prepared after the collection of three spectra.

**III. Results and Discussion**

LEIS data are normally analyzed classically and with the binary collision approximation (BCA) [20] due to the small de Broglie wavelength of the incoming ions and the small ratio between the scattering cross section and the interatomic spacings, respectively. In the BCA, the scattering process is considered as a series of interactions of the projectile with one atom in the crystal at a time. The ions that experience only one hard collision with a surface atom that is visible to both the incoming ion beam and the detector produce a sharp single scattering peak (SSP) in a LEIS spectrum. The kinetic energy of the scattered projectiles in a SSP are determined primarily by the projectile/target mass ratio and the scattering angle, basically providing a mass spectrum of the surface composition [15]. To study defected and oxygen exposed Gr/Ru(0001), the areas of the C, O and Ru SSPs are computed by integration after subtracting the baseline of multiply scattered ions, which is modeled via a polynomial fit of the shoulder of the SSPs. When computing the ratio of SSP areas between different elements, a normalization based on the relative differential cross sections is employed [15,21].

In order to backscatter from light target atoms, such as C and O, a light projectile is required, such as helium. Another advantage of using $He^+$ for ion scattering is its high surface sensitivity due to Auger neutralization (AN) [15,22]. In AN, most of the projectiles that collide with multiple target atoms are neutralized due to the relatively large He ionization level compared to the size of the surface conduction band. Thus, the signal captured by the ESA consists primarily of single scattering events from the outermost atomic layers. In addition, there is a strong matrix effect when helium ions are scattered from graphitic carbon [16,23,24]. This effect involves a quasi-resonant neutralization in conjunction with AN, which leads to a very high neutral fraction for primary incident beam energies below 2500 eV [25]. Therefore, to avoid the matrix effect and provide a strong signal of scattered $He^+$, an incident beam energy of 3000 eV is used here.

The experiments are performed using a specular geometry in which the incident and exit angles are equal with respect to the surface normal, and a scattering of angle or either 45° or 115° is employed [26]. At 45°, the incident and exit angles are grazing enough that only the outermost Gr layer and any adatoms adsorbed on it are probed. At the larger 115° angle, the ions penetrate the overlayer so that the spectra also probe any intercalates and some of the underlying substrate.

The cleanliness of the Ru(0001) surface and the growth of graphene onto it are verified with He$^+$ LEIS and LEED. The clean Ru(0001) surface shows only a Ru SSP in LEIS spectra and a 1x1 LEED pattern. After producing a clean surface, LEIS at a scattering angle of 45° is used to monitor the outermost surface after the growth of graphene. Spectrum (a) in Fig. 1 shows only a C SSP, which indicates that the graphene overlayer completely covers the substrate. Since the cross section for the scattering from Ru is about 16 times larger than for scattering from C, this is a very sensitive measurement of the completeness of the overlayer. In addition, LEED images collected from this surface show a Moiré pattern due to the formation of a superlattice between the Gr film and the Ru(0001) substrate, as reported previously [27,28].

**A. The role of defects in intercalation**

The defects on Gr/Ru(0001) are introduced by a 3 min 50 eV Ar$^+$ sputtering, and 3 keV He$^+$ ion scattering spectra collected at a 45° scattering angle are used to monitor any changes to the surface, as shown in spectrum (b) of Fig. 1. The sharp C SSP is still clearly seen in LEIS and the LEED Moiré pattern remains sharp, which indicate that the structure of the graphene overlayer is not overtly affected by the light sputtering. There is a small Ru SSP in the spectrum, which indicates that a small number of C vacancy defects are introduced by the light sputtering that reveal a small part of the substrate. This low fluence of Ar$^+$ sputtering produces isolated single carbon vacancy defects, as described elsewhere [26].

Because the sample is annealed when investigating the intercalation of oxygen and the reaction of intercalated oxygen with graphene below, the sputtered surface is annealed here as a control prior to oxygen exposure. The LEIS spectrum collected after the lightly sputtered sample is annealed to 1000 K is shown in spectrum (c) of Fig. 1. It is seen that the areas of the C and Ru

SSPs do not change and the Moiré LEED pattern is retained after the annealing, which implies that the defects on Gr/Ru(0001) are themselves stable, at least for annealing under 1000 K.

In addition, as shown in our recent work [26], oxygen molecules can physically absorb onto isolated carbon vacancy defects, even at room temperature, and the adsorbed oxygen diffuses and intercalates between graphene and Ru(0001) when the sample is heated to 600 K. To avoid the influence of oxygen attached to defect sites in the present measurements, the samples are held at 650 K during the $O_2$ exposures so that the all of the molecules intercalate. To confirm the location of oxygen after exposure of defective Gr/Ru(0001), a $He^+$ ion scattering spectrum is collected at 45° after a 3 min sputtering followed by a 6400 L $O_2$ exposure at 650 K, as shown in Fig. 2. In this case, only the C SSP and a small Ru SSP are observed, which indicates that any adsorbed oxygen is intercalated below the Gr overlayer and none is present atop the Gr or at defect sites following $O_2$ exposure at elevated temperature.

To explore how the oxygen intercalation efficiency is affected by the isolated carbon vacancies produced by a light pre-sputtering, an $O_2$ exposure of 900 L at 650 K is performed on the as-prepared Gr/Ru(0001) and on Gr/Ru(0001) with 1 min pre-sputtering and 3 min pre-sputtering, and the subsequent LEIS spectra are shown in Fig. 3. During the exposure, the sample is held at 650 K, and the spectra are collected at a 115° scattering angle to probe beneath the Gr film after the sample has cooled to room temperature. The figure shows that the areas of C SSPs do not change, but the areas of O SSPs increase with sputtering time. It is thus inferred that the oxygen intercalates easier on Gr/Ru(0001) as the concentration of C vacancy defects increases.

In fact, it is shown in the LEEM images of Ref. [10] that the oxygen intercalates from the edges of graphene flakes on a Ru substrate. This suggests that it is energetically favorable for oxygen to adsorb at the edge of the graphene film and to then diffuse to become intercalated if the surface temperature is high enough to enable the diffusion. The mechanism is very similar in the presence of defects. As proposed in previous reports [29-33], the defects and dopants on 2D materials, such as $MoS_2$, boron nitride or graphene, act as active sites for molecular adsorption even at room temperature. In particular, our previous study of defects on Gr/Ru(0001) [26] showed that the oxygen adsorbs molecularly, even at room temperature, on single vacancy defects created by light sputtering and then diffuses between the graphene film and Ru(0001) substrate when annealed to 600 K. It is thus reasonable to assume that when defect containing Gr/Ru(0001) is exposed to $O_2$ at elevated temperature, the defect sites behave as transition stations for the intercalation process, in which they keep adsorbing and transferring the adsorbates to the more stable sites beneath the graphene overlayer. This process increases the intercalation rate as compared to that of the original intact graphene layer. This indicates that defects further degrade the quality of graphene films on metal substrates, beyond the presence of the defects themselves, by increasing the probability and rate of intercalation. This effect will be particularly critical in applications in which graphene materials are exposed to an ambient environment and/or prepared at an elevated temperature.

**B. The role of defects in the etching reaction**

It has been reported that oxygen intercalated between Gr and Ru(0001) starts to desorb from the sample and react with the graphene overlayer when annealed to 750 K or above [13,34]. The annealing leads to both the removal of molecular $O_2$ and the etching of some of the graphene, presumably through the formation of gaseous CO or $CO_2$.

To explore how this etching is affected by carbon vacancy defects, spectra collected from the as-prepared Gr/Ru(0001) are compared to those collected from the defect-containing samples in Fig. 4. The as-prepared Gr/Ru(0001) samples and those pre-sputtered for 1 min and 3 min are exposed to 6400 L of $O_2$ while being held at 650 K. Following the $O_2$ exposures, an additional 10 min post-annealing at 1000 K is employed to force the complete desorption of the intercalated oxygen. Each panel of Fig. 4 shows $He^+$ ion scattering spectra collected from the $O_2$ exposed samples both before (dashed lines) and after (solid lines) the post-annealing. The spectra are all collected at 115°, so that the intercalated oxygen is detected. It is seen in Fig. 4 that the oxygen SSPs are absent after the annealing, confirming that the desorption process is complete after annealing to 1000 K. The etching of graphene during the desorption process is indicated by the decrease of the C SSPs following annealing. It is clearly seen that magnitude of the decrease of the C SSP area increases with the fluence of the pre-sputtering treatment.

In Fig. 5, the absolute amounts of C in the graphene film and intercalated O before and after the annealing are plotted, along with the ratio of C to the O decrease following the desorption process, as a function of sputtering time. The amount of C was calculated from the LEIS data shown in Fig. 4 by assuming that the as-grown sample represents a full ML of Gr, and that the C SSP area is proportional to the Gr coverage. The oxygen coverage was calculated by normalizing the O SSP area to that of the C SSP after correcting for the differential cross section. Note that the O coverage may be an underestimate, as the He ions scattered from intercalated oxygen may be more prone to neutralization than those scattered from the Gr film, and some of oxygen may also be shadowed by the surface Gr. Nevertheless, the data in the figure does correctly show the trends in how the oxygen coverage and the ratio of lost C to lost O change.

It is clearly seen that the amount of oxygen decreases to the same level after the post-annealing for all sputtering times. Since the intercalation efficiency depends on the number of defects on the surface, as discussed above, it is reasonable that a longer pre-sputtering time leads to a slightly larger amount of oxygen intercalation even though the exposures are the same, which can be seen in the data of Fig. 5. Thus, the decrease of the amount of intercalated oxygen during the desorption process increases with pre-sputtering time because there is more oxygen that can be removed.

Meanwhile, the decrease in the amount of surface carbon following annealing is much more affected by the number of surface defects. Although it may be expected that the C coverage would not decrease in the absence of defects, the data show a loss of 0.066 ML of C when the as-prepared sample is heated, as was reported previously [13]. The data further show, however, that with increased sputtering time, the absolute decrease in the amount of C and the ratio of lost C to lost O during desorption both increase. That indicates that the defects introduced by the sputtering treatment improve the etching efficiency of the graphene layer.

A reasonable explanation for how the defects created by the pre-sputtering ease the etching of graphene by intercalated oxygen is that the C atoms in the vicinity of the defects are more active in contrast to their inert behavior when intercalated beneath a complete graphene layer. As reported in Ref. [14], single carbon vacancies are introduced on Gr/Pt(111) by 140 eV $Ar^+$ sputtering, which is also a weak graphene-metal interacting system similar to Gr/Ru(0001). In their DFT simulations, a 6×6 unit cell (4 cells of the 3×3 Gr/Pt(111) Moiré) was used. It was found that two of the three undercoordinated C atoms that surround a single C vacancy move closer to each other and form a pentagonal ring, becoming weakly bonded so that they could be etched first during the desorption process. In addition, the third atom and one of its neighbors moves out of the initial graphene plane towards the Pt substrate (distance decreases from 3.27 Å to 1.97 Å), forming two new chemical bonds with Pt atoms. This likely further improves the etching efficiency as the carbon-metal bond is easier to break than the carbon-carbon bond in defect-free graphene. It is also noted that the average distance between the graphene overlayer and the Pt(111) substrate decreases by 0.1 Å due to the presence of the C vacancies, which could also enhance the etching reaction at the edges of the Gr films.

## IV. Conclusions

Small molecules, such as $O_2$, can intercalate between a graphene film and the substrate and the intercalated oxygen will etch the graphene overlayer when the sample is annealed to remove the intercalated oxygen [9,10,13,34]. To further study the factors that affect the oxygen intercalation and the etching of Gr during the thermal desorption of oxygen, a light sputtering of the Gr/Ru(0001) surface was performed before the $O_2$ exposures to create single carbon vacancy defects. The data show that those defects improve the efficiency of oxygen intercalation. This implies that it is necessary to maintain the completeness of a graphene overlayer when storing graphene-based devices, as defects will cause the devices to degrade more quickly, especially when heating or fabricating them in air or low vacuum. Furthermore, it is shown that etching of the Gr film during the desorption of intercalated oxygen is more efficient in the presence of defects. It is thus concluded that the C atoms located near the defects or edges of the Gr film are less stably bonded and can thus more easily react with intercalated oxygen to form the gaseous etching products.

## V. Acknowledgements

This material is based upon work supported by the National Science Foundation under CHE - 1611563.

# References

[1] K. S. Novoselov, A. K. Geim, S. V. Morozov, D. Jiang, M. I. Katsnelson, I. V. Grigorieva, S. V. Dubonos, and A. A. Firsov, Nature **438**, 197 (2005).

[2] C. Berger, Z. Song, X. Li, X. Wu, N. Brown, C. Naud, D. Mayou, T. Li, J. Hass, A. N. Marchenkov, E. H. Conrad, P. N. First, and W. A. de Heer, Science **312**, 1191 (2006).

[3] A. K. Geim, Science **324**, 1530 (2009).

[4] K. S. Kim, Y. Zhao, H. Jang, S. Y. Lee, J. M. Kim, K. S. Kim, J.-H. Ahn, P. Kim, J.-Y. Choi, and B. H. Hong, Nature **457**, 706 (2009).

[5] X. Li, W. Cai, J. An, S. Kim, J. Nah, D. Yang, R. Piner, A. Velamakanni, I. Jung, E. Tutuc, S. K. Banerjee, L. Colombo, and R. S. Ruoff, Science **324**, 1312 (2009).

[6] P. W. Sutter, J.-I. Flege, and E. A. Sutter, Nature Materials **7**, 406 (2008).

[7] D. Kang, J. Y. Kwon, H. Cho, J.-H. Sim, H. S. Hwang, C. S. Kim, Y. J. Kim, R. S. Ruoff, and H. S. Shin, ACS Nano **6**, 7763 (2012).

[8] D. Prasai, J. C. Tuberquia, R. R. Harl, G. K. Jennings, and K. I. Bolotin, ACS Nano **6**, 1102 (2012).

[9] R. Larciprete, S. Ulstrup, P. Lacovig, M. Dalmiglio, M. Bianchi, F. Mazzola, L. Hornekær, F. Orlando, A. Baraldi, P. Hofmann, and S. Lizzit, ACS Nano **6**, 9551 (2012).

[10] P. Sutter, J. T. Sadowski, and E. A. Sutter, J. Am. Chem. Soc. **132**, 8175 (2010).

[11] D. Ma, Y. Zhang, M. Liu, Q. Ji, T. Gao, Y. Zhang, and Z. Liu, Nano Research **6**, 671 (2013).

[12] X. Feng, S. Maier, and M. Salmeron, J. Am. Chem. Soc. **134**, 5662 (2012).

[13] T. Li and J. A. Yarmoff, Phys. Rev. B **96**, 155441 (2017).

[14] M. M. Ugeda, D. Fernández-Torre, I. Brihuega, P. Pou, A. J. Martínez-Galera, R. Pérez, and J. M. Gómez-Rodríguez, Phys. Rev. Lett. **107**, 116803 (2011).

[15] W. J. Rabalais, *Principles and applications of ion scattering spectrometry : surface chemical and structural analysis* (Wiley, New York, 2003).

[16] S. Prusa, P. Prochazka, P. Babor, T. Sikola, R. ter Veen, M. Fartmann, T. Grehl, P. Bruner, D. Roth, P. Bauer, and H. H. Brongersma, Langmuir **31**, 9628 (2015).

[17] H. Zhang, Q. Fu, Y. Cui, D. Tan, and X. Bao, J. Phys. Chem. C **113**, 8296 (2009).

[18] Z. Zhou, F. Gao, and D. W. Goodman, Surf. Sci. **604**, L31 (2010).

[19] M. Pozzo, D. Alfè, P. Lacovig, P. Hofmann, S. Lizzit, and A. Baraldi, Phys. Rev. Lett. **106**, 135501 (2011).

[20] M. T. Robinson and I. M. Torrens, Phys. Rev. B **9**, 5008 (1974).

[21] H. Niehus, W. Heiland, and E. Taglauer, Surf. Sci. Rep. **17**, 213 (1993).

[22] H. H. Brongersma, M. Draxler, M. de Ridder, and P. Bauer, Surf. Sci. Rep. **62**, 63 (2007).

[23] S. N. Mikhailov, L. C. A. van den Oetelaar, and H. H. Brongersma, Nucl. Instrum. Methods Phys. Res., Sect. B **93**, 210 (1994).

[24] L. C. A. van den Oetelaar, S. N. Mikhailov, and H. H. Brongersma, Nucl. Instrum. Methods Phys. Res., Sect. B **85**, 420 (1994).

[25] J. B. Hasted, *Physics of Atomic Collisions* (Butterworths, London, 1972).

[26] T. Li and J. A. Yarmoff, to be published.

[27] A. B. Preobrajenski, M. L. Ng, A. S. Vinogradov, and N. Mårtensson, Phys. Rev. B **78** (2008).

[28] B. Wang, M. L. Bocquet, S. Marchini, S. Gunther, and J. Wintterlin, Phys. Chem. Chem. Phys. **10**, 3530 (2008).

[29] B. Zhao, C. Shang, N. Qi, Z. Y. Chen, and Z. Q. Chen, Appl. Surf. Sci. **412**, 385 (2017).

[30] Z.-Y. Z. Deng, Jian-Min, Can. J. Phys. **94**, 10 (2016).

[31] Q. Zhou, W. Ju, Y. Yong, X. Su, X. Li, Z. Fu, and C. Wang, Physica E **95**, 16 (2018).

[32] B. Sanyal, O. Eriksson, U. Jansson, and H. Grennberg, Phys. Rev. B **79**, 113409 (2009).

[33] H. Guang, M. Aoki, S. Tanaka, and M. Kohyama, Solid State Commun. **174**, 10 (2013).

[34] A. Dong, Q. Fu, M. Wei, Y. Liu, Y. Ning, F. Yang, H. Bluhm, and X. Bao, Surf. Sci. **634**, 37 (2015).

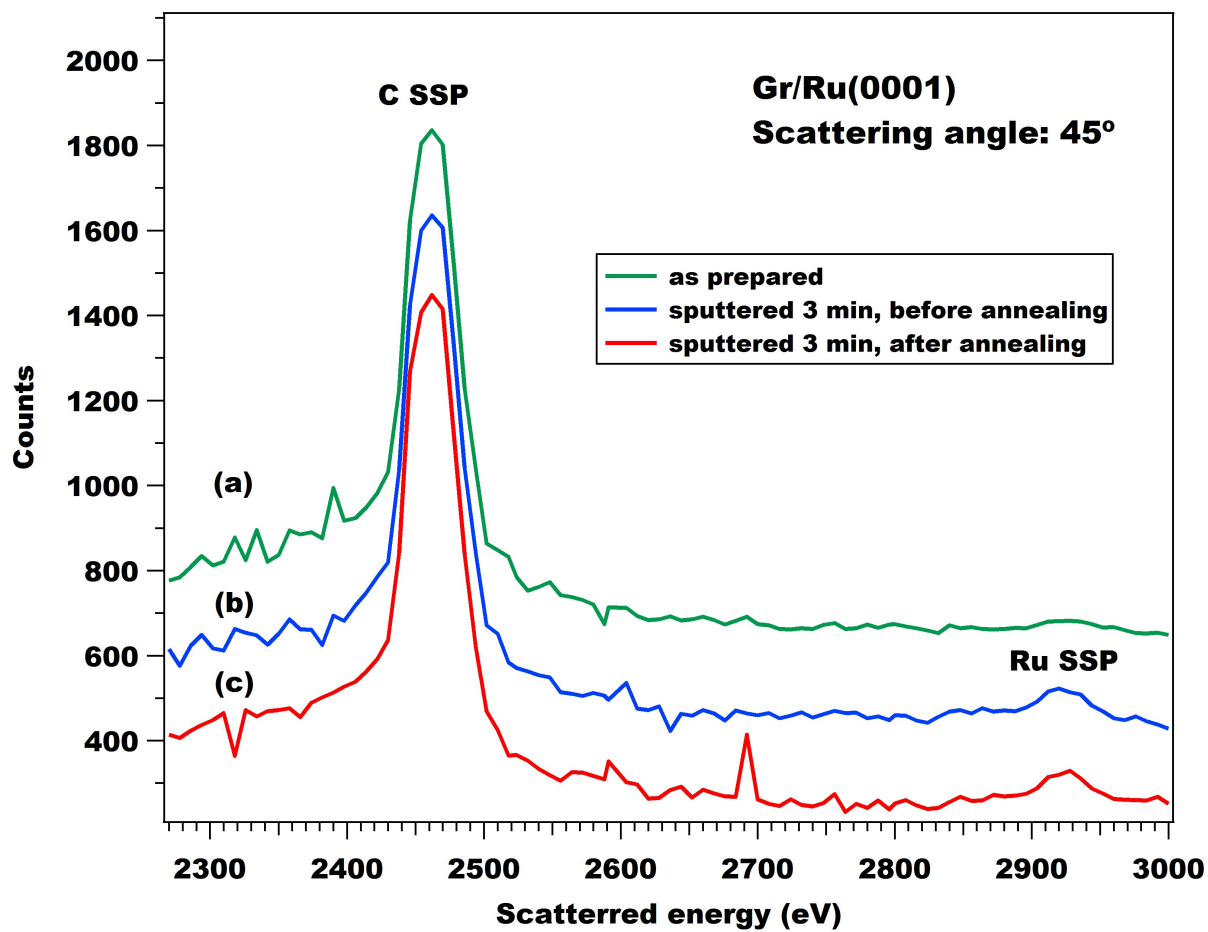

**Figure 1**. 3.0 keV He$^+$ ion scattering spectra collected at a 45° scattering angle from (a) as prepared Gr/Ru(0001), (b) after a 3 min light sputtering, and (c) after post-annealing at 1000 K. The y-axes of the spectra are offset for clarity.

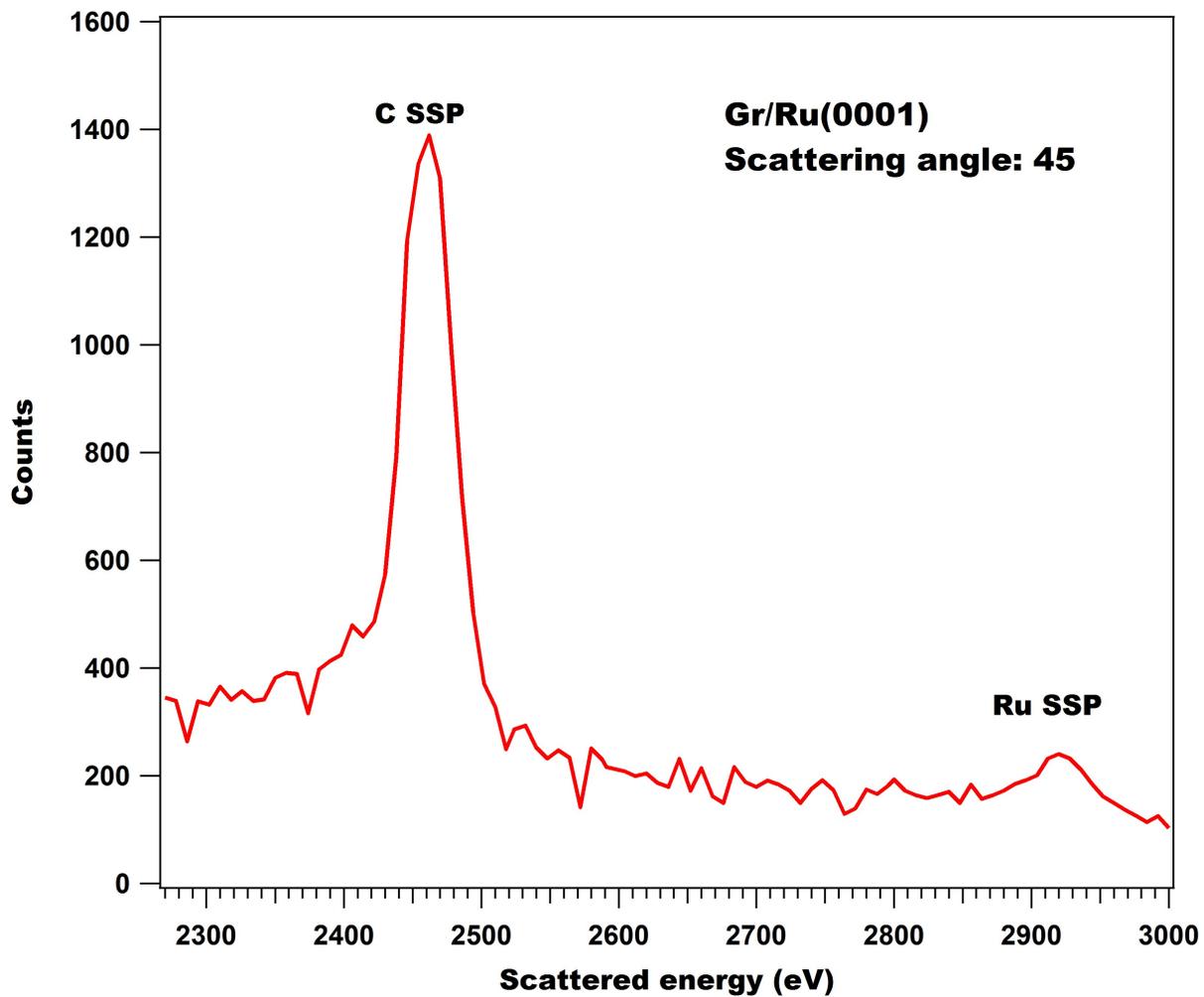

**Figure 2.** 3.0 keV He$^+$ ion scattering spectrum collected at a 45° scattering angle from Gr/Ru(0001) after a 3 min light sputtering followed by exposure to 6400 L of O$_2$ at 650 K.

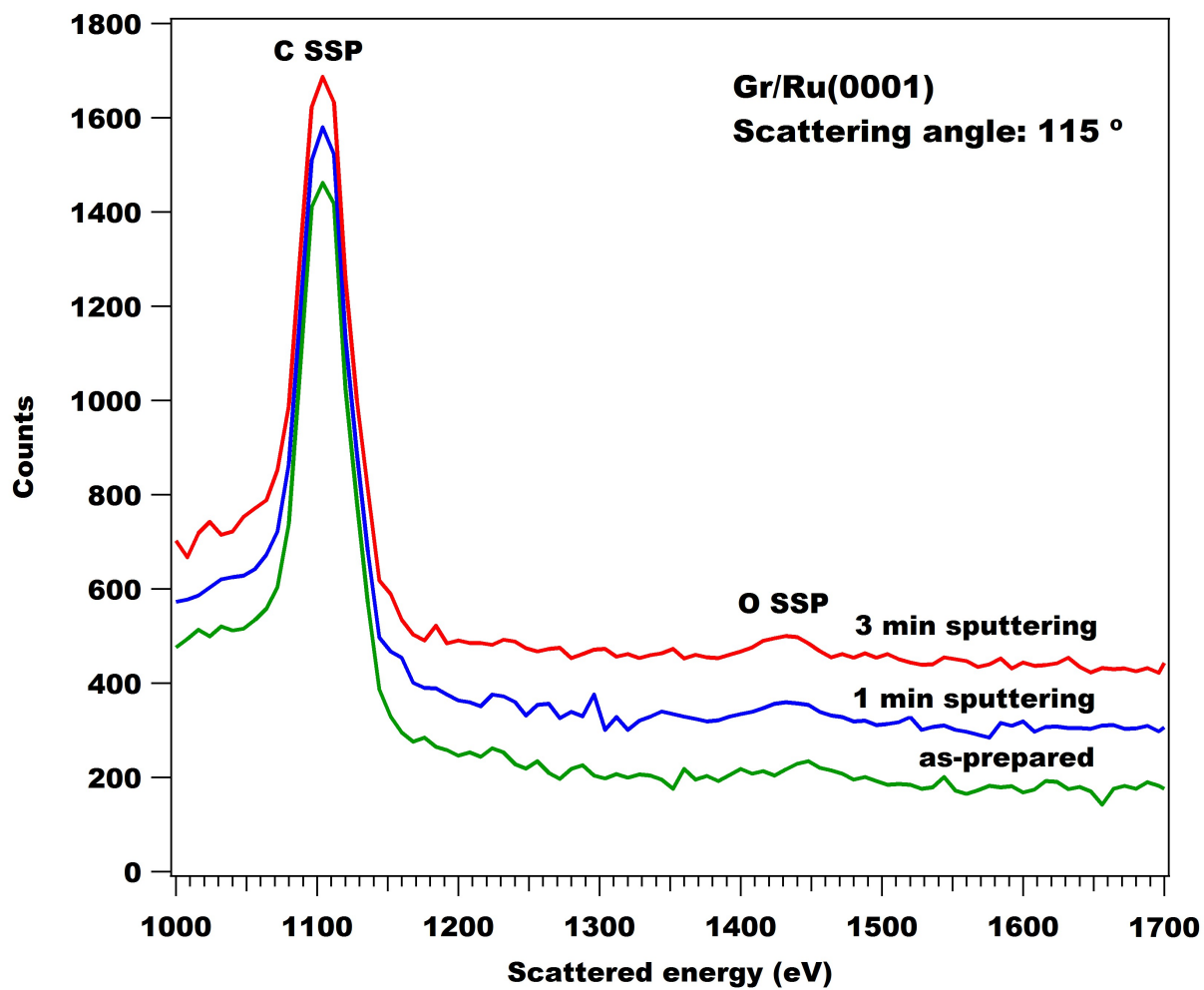

**Figure 3.** 3.0 keV He$^+$ LEIS spectra collected at a scattering angle of 115° from Gr/Ru(0001) exposed to 900 L of O$_2$ at 650 K following the indicated amounts of pre-sputtering. The spectra were collected after the sample cooled to room temperature.

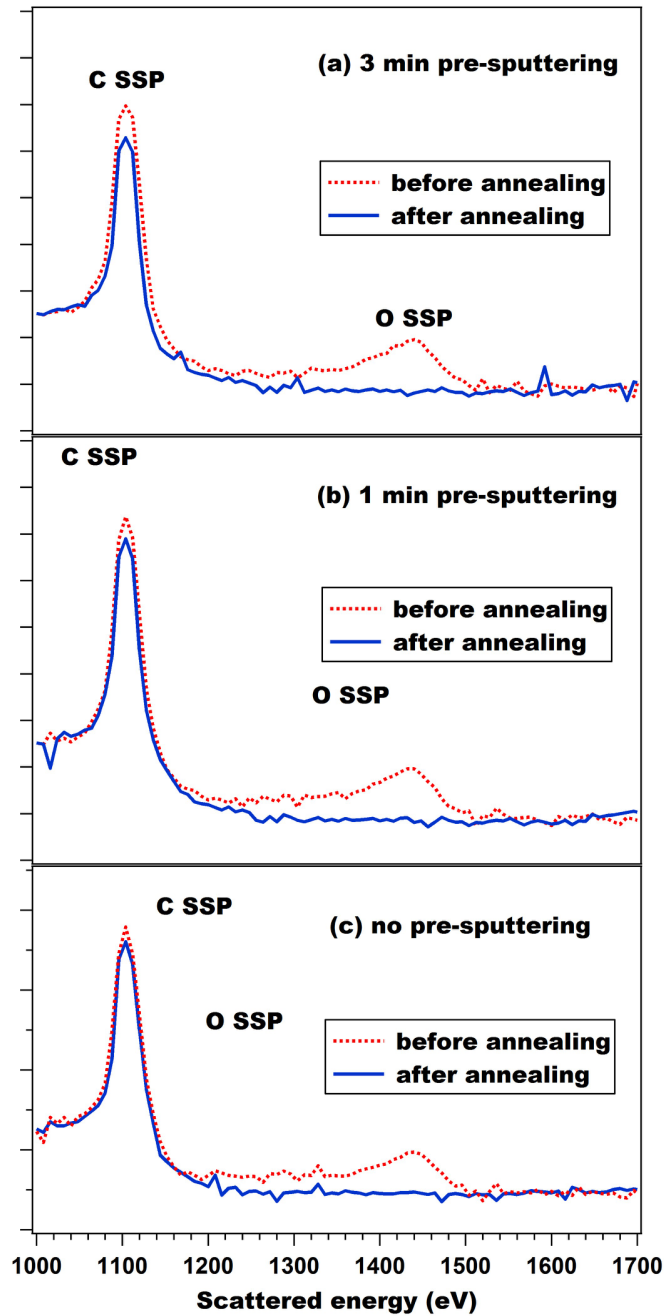

**Figure 4.** 3.0 keV He$^+$ LEIS spectra collected at a scattering angle of 115° from Gr/Ru(0001) exposed to 6400 L of O$_2$ at 650 K (a) after a 3 min sputtering, (b) after a 1 min sputtering, and (c) without sputtering. The dashed lines show spectra collected right after the oxygen exposure, while the solid lines show spectra collected after an additional annealing at 1000 K for 10 min.

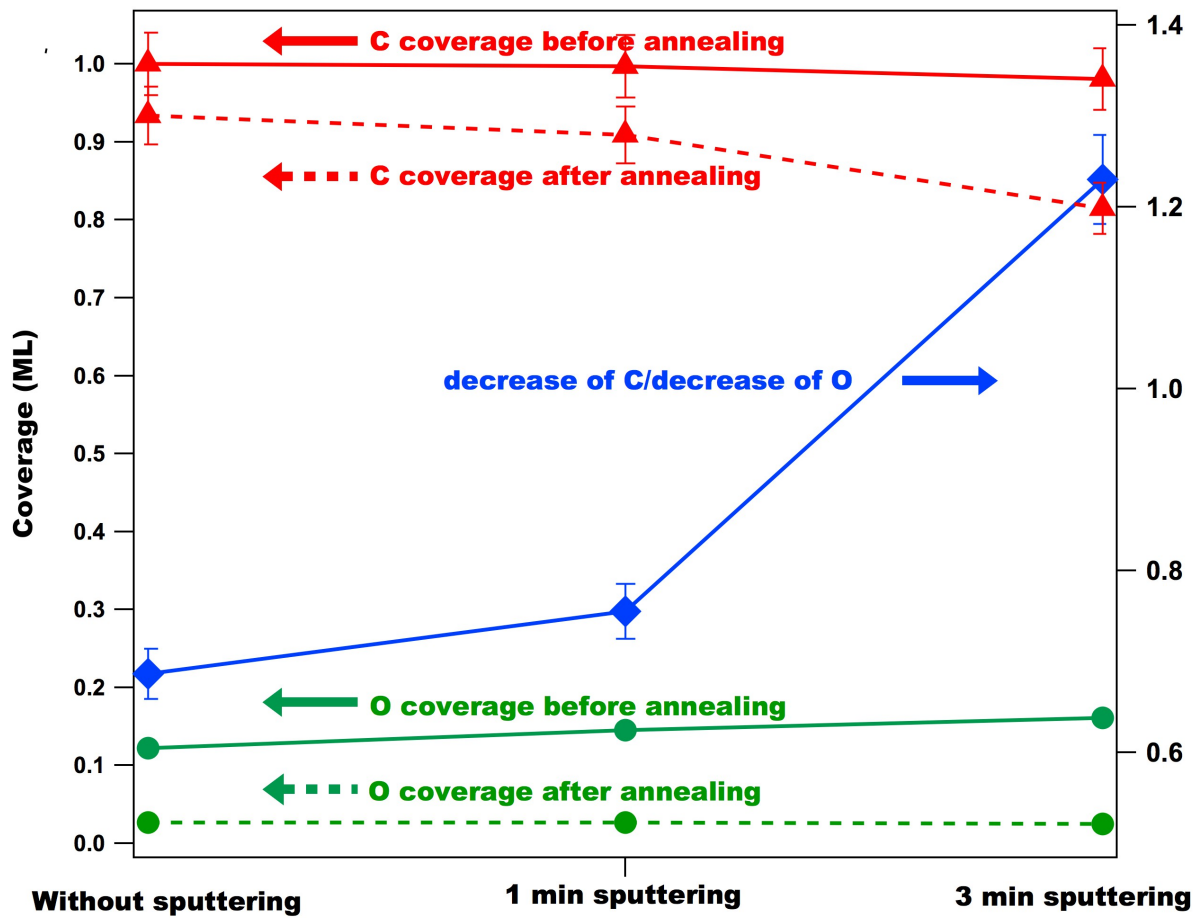

**Figure 5.** The left axis is used to show the amount of C and O on the surface before and after a 10 min annealing at 1000 K, based on the data from Fig. 4, and are shown as a function of the amount of pre-sputtering before the $O_2$ exposure. The right axis is used to indicate the ratio of the decrease of C coverage to the decrease of O coverage during the oxygen desorption. The ratio calculation includes a normalization of the SSP areas by the respective differential cross sections for He scattering from C and O.